\documentclass[aps,pre,floats,floatfix,twocolumn]{revtex4-1}

\usepackage{ifthen}
\usepackage{ifpdf}
\usepackage{color}

\ifpdf
\usepackage{graphicx}
\usepackage{epstopdf}
\else
\usepackage{graphicx}
\usepackage{epsfig}
\fi
\graphicspath{{./Figs/}}

\usepackage{latexsym}
\usepackage{amsmath}
\usepackage{amssymb}
\usepackage{bm}
\usepackage{wasysym}

\DeclareSymbolFont{epsilon}{OML}{ntxmi}{m}{it}
\DeclareMathSymbol{\epsilon}{\mathord}{epsilon}{"0F}

\usepackage{hyperref}

\newcommand{\im}{\mbox{Im}}
\newcommand{\re}{\mbox{Re}}

\newcommand{\amatrix}[1]{\begin{matrix} #1 \end{matrix}} 
\newcommand{\eexp}[1]{\mathrm{e}^{#1}}

\newcommand{\braket}[1]{ \left\langle #1 \right\rangle}

\newcommand{\ola}{\protect\overleftarrow}
\newcommand{\ora}{\protect\overrightarrow}

\newcommand{\be}[1]{\begin{eqnarray}{\label{e#1}}} 
\newcommand{\beq}{\begin{eqnarray}}
\newcommand{\eeq}{\end{eqnarray}} 

\newcommand{\hide}[1]{}

\newcommand{\Eq}[1]{\textcolor{blue}{{Eq.}\!~(\ref{#1})}}

\newcommand{\Fig}[1]{\textcolor{blue}{Fig.}\!\!~\ref{#1}}

\newcommand{\rmrkold}[1]{{#1}}     
\newcommand{\rmrk}[1]{{#1}}        
\newcommand{\hrefl}[1]    {\href{#1}{[link]}}
\newcommand{\hidea}[1]{}

\renewcommand{\url}[2] {}   
\newcommand{\urlprefix}{}

\begin{document}
 
\title{The relaxation rate of a stochastic spreading process in a closed ring}

\author{Daniel Hurowitz, Doron Cohen}

\affiliation{Department of Physics, Ben-Gurion University of the Negev, Beer-Sheva, Israel}

\begin{abstract}
The relaxation process of a diffusive ring becomes under-damped if the bias (so called affinity) exceeds a critical threshold value, \rmrkold{aka delocalization transition}. 
This is related to the spectral properties of the pertinent stochastic kernel. We find the dependence of the relaxation rate on the affinity and on the length of the ring. Additionally we study the implications of introducing a weak-link into the circuit, and illuminate some subtleties that arise while taking the continuum limit of the discrete model.
\end{abstract}

\maketitle


\section{Introduction}

In the absence of topology the relaxation time of a stochastic sample is
\rmrk{determined either by the diffusion or by the drift,} 
depending on whether the bias is small or large, respectively. 
\rmrk{In contrast,} in a topologically closed circuit, as the bias is increased, 
\rmrk{the relaxation becomes under-damped  
with relaxation-rate that is determined by the diffusion and not by the drift}. 
In related applications the ``circuit" might be a chemical-cycle, 
and the ``bias" is the so called {\em affinity} of the cycle. 

In the present work we consider a minimal model for a topologically
closed circuit, namely, an $N$~site ring with nearest-neighbor hopping.
The dynamics can be regarded as a stochastic process 
in which a particle hops from site to site. 
The rate equation for the site \rmrk{occupation probabilities $\bm{p}  = \{p_n\}$}
can be written in matrix notation as 
\be{1}
\frac{d\bm{p}}{dt} \ \ = \ \ \bm{W} \bm{p}, 
\eeq
If the ring were opened, then the $N\to\infty$ limit would correspond 
to Sinai's spreading problem \cite{Sinai,odh1,odh3,BouchaudReview}, 
aka {\em a random walk in a random environment}, 
where the transition rates are allowed to be asymmetric. 
Such models have diverse applications, 
notably in biophysical contexts of populations biology \cite{popbio,popbio2} pulling pinned polymers and DNA unzipping \cite{DNA1,DNA2}
and in particular with regard to  molecular motors~\cite{fisher1999force,rief2000myosin,brm1,brm2}. 

In the absence of topology $\bm{W}$ is {\em similar} to a real symmetric matrix, 
and the relaxation spectrum is {\em real} \rmrk{(aka damped relaxation)}. 
Alas, for a ring the affinity is a topological-invariant 
that cannot be gauged away, analogous to the Aharonov-Bohm flux, 
\rmrk{and the relaxation spectrum might become {\em complex} (aka under-damped relaxation).}  
Thus the theme that we are addressing here is related to the study 
of of non-Hermitian quantum Hamiltonians \cite{Hatano1,Hatano2,Shnerb1}. 
In a previous work \cite{neg} we have illuminated the relation between
the sliding-transition and the complexity-threshold, 
aka ``de-localization transition", as the affinity is increased.

The outline is as follows: 
In Sec.II we discuss the relaxation in the case of an homogeneously 
disordered diffusive sample, contrasting non-trivial topology (ring) 
with simple geometry (box).  
The effect of disorder is demonstrated in Sec.III, 
\rmrkold{where} heuristic considerations are used in order 
to explain the dependence of the relaxation rate 
on the affinity and on the length of the ring.  
In Sec.IV we discuss the delocalization transition.
Namely, we find the threshold value of the affinity 
beyond which the relaxation becomes under-damped. 
Then we extract the relaxation rate from the characteristic 
equation using an ``electrostatic picture".
\rmrkold{As explained in Section V the same picture 
can be used in order to address sparse disorder. 
This motivates the analysis in Sections VI-VII} 
of the relaxation in a ring that has an additional weak-link 
that forms a bottleneck for diffusion, 
though not blocking it completely.
Several appendices are provided to make the presentation self-contained.

\section{Diffusive sample: Ring vs Box}

The rate equation \Eq{e1} involves a matrix ${\bm{W}}$ 
whose off-diagonal elements are the transition rates ${w_{nm}}$, 
and whose diagonal elements are ${-\gamma_n}$ such that each column sums to zero.
\rmrkold{Via diagonalization one can find the eigenvalues ${\{-\lambda_{\nu}\}}$. 
Irrespective of models details there always exists an eigenvalue ${\lambda_0=0}$ 
that corresponds to the non-equilibrium steady state (NESS). 
The other eigenvalues reflect the relaxation modes of the system:   
they have positive~$\re[\lambda_{\nu}]$, and might be complex.
Complexity of the low eigenvalues implies an under-damped relaxation.} 

For a clean ring and with near-neighbor hopping, 
the rates are uniform but asymmetric, 
and are equal to $\ora{w}=we^{s/2}$ for forward hopping, 
and  $\ola{w}=we^{-s/2}$ for backward hopping. 
The ${\bm{W}}$ matrix takes the form
\be{2}
\bm{W} \ \ = \ \ \left[\amatrix{
-\gamma     & \ola{w}   & 0         & ...  & \ora{w} \\ 
\ora{w}     & -\gamma   & \ola{w}   & ...  & ... \\ 
0           & \ora{w}   & -\gamma   & ...  & ... \\
...         & ...       & ...       & ...  & ... \\
\ola{w}     & ...       & ...       & ...  & ...
}\right]
\eeq 
with $\gamma = -2w\cosh(s/2)$. 
Due to translational invariance, 
this matrix can be written in terms of momentum operator
\be{3}
{\bm W} = we^{s/2 + i\bm{P}}  + we^{-s/2-i\bm{P}}  - 2w\cosh\left(\frac{s}{2}\right)
\eeq
From here it is easy to see that the eigenvalues are  
\be{12}
\lambda_{\nu} = 2w \ \left[ \cosh\left(\frac{s}{2}\right)-\cos\left(\frac{2\pi}{N}\nu + i\frac{s}{2}\right)\right]
\eeq
The complexity of the $\nu\neq0$ eigenvalues implies 
that the relaxation process in not over-damped. 
A straightforward analysis of the time-dependent spreading process, see e.g. \cite{nes}, 
shows that the drift velocity and the diffusion coefficient are given by the following expressions:
\beq
v_0 &=& \ (\ora{w}-\ola{w})a \ \ = \ 2wa \ \sinh(s/2) 
\\ \label{e106}
D_0 &=& \frac{1}{2}(\ora{w}+\ola{w})a^2 \ = \ wa^2 \ \cosh(s/2)
\eeq  
where $a$ is the lattice constant. 
Note that in \Eq{e3} we used the lattice constant 
as a unit of length (``${a=1}$") else the following 
replacement is required: $\bm{P}\mapsto a\bm{P}$.

It is convenient to consider the continuum limit of the rate equation \Eq{e1}.
In this limit we define ${D(x)=wa^2}$ and ${v(x)=swa}$,   
and the continuity equation for the \rmrk{probability density $\rho(x_n)=(1/a)p_n$}  
becomes the Fokker-Planck diffusion equation:
\be{107}
\frac{d\rho}{dt} \ \ = \ \ -\frac{d}{dx}\left[ -D(x)\frac{d\rho}{dx} + v(x) \rho(x) \right] 
\eeq 
One can easily find the spectrum of the relaxation modes \rmrkold{(${\re[\lambda_{\nu}]>0}$)} 
for either ``ring" or ``box" geometry.
\rmrk{The length of the segment is ${L=Na}$,} and the boundary conditions 
are respectively either of Neumann type or periodic. The result is 
\be{121}
\lambda_{\nu} \mbox{[ring]} & \ = \ & \left(\frac{2\pi}{L}\right)^2D  \nu^2 \ +\ i \frac{2\pi v}{L}\nu
\\ 
\lambda_{\nu} \mbox{[box]} & \ = \ &  \left(\frac{\pi}{L}\right)^2D  \nu^2 \ + \ \frac{v^2}{4D} 
\eeq
where for the ring ${\nu=\pm1,\pm2,... }$,  
while for the box ${\nu=1,2,3,...}$. 
Clearly \Eq{e121} is consistent with \Eq{e12}.
The relaxation rate $\Gamma$ is determined by the lowest eigenvalue
\be{564}
\Gamma  \ \ \equiv \ \ \re[\lambda_1] 
\eeq
\rmrk{For the ``ring" it is determined solely by the diffusion coefficient:}
\be{13}
\Gamma \mbox{[ring]} \ \ = \ \ \left(\frac{2\pi}{L}\right)^2 D 
\eeq   
while for the ``box", \rmrk{if the bias is large, it is predominantly determined by the drift:} 
\be{14}
\Gamma \mbox{[box]} \ \ = \ \ \left[\left(\frac{\pi}{L}\right)^2  + \left(\frac{v}{2D}\right)^2\right]D 
\eeq  
It is important to realize that in the latter case we 
have a ``gap" in the spectrum, meaning that $\lambda_1$  
does not diminish in the ${L\rightarrow \infty}$ limit, 
hence the relaxation time is finite.

\section{Disordered ring}

\rmrkold{In the presence of disorder, the forward and backward rates 
across the $n$th bond are random numbers $\ora{w}_n$ and~$\ola{w}_n$. 
Accordingly the diagonal elements of $\bm{W}$ 
are random too, namely ${\gamma_n= \ola{w}_n + \ora{w}_{n{+}1}}$.}
By considering the long time limit of the time-dependent spreading process
it is still possible to define the drift velocity~$v$ and diffusion coefficient~$D$.   
The results depend in an essential way on the affinity of the cycle 
\beq
S_{\circlearrowleft} \ \ \equiv \ \ N \ s 
\eeq 
where $s$ is defined via the sample average 
\beq
\rmrkold{ \frac{1}{N} \sum_{n=1}^{N}  \ln\left(\frac{\ola{w}_n}{\ora{w}_n}\right) } \ \ \equiv \ \ -s
\eeq
Additionally it is useful to define threshold values $s_{\mu}$,
\rmrkold{whose significance will be clarified in the next section}, 
via the following expression:
\be{362}
\rmrkold{ \frac{1}{N} \sum_{n=1}^{N} \left(\frac{\ola{w}_n}{\ora{w}_n}\right)^{\mu} }  \ \ \equiv \ \ \eexp{-(s-s_{\mu})\mu} 
\eeq
Here, as in \cite{nes,neg} we assume that the rates are 
\beq
\ora{w}_n \ &=& \ w \ \eexp{+\mathcal{E}_n/2} \\
\ola{w}_n \ &=& \ w \ \eexp{-\mathcal{E}_n/2} 
\eeq
where the ``activation energies" $\mathcal{E}_n$ 
are box distributed within ${[s-\sigma,s+\sigma]}$. 
\rmrkold{Approximating the sample average by 
an ensemble average} the thresholds of \Eq{e362} are 
\be{15}
s_{\mu} \ \ = \ \ \frac{1}{\mu} \ln\left( \frac{\sinh (\sigma\mu)}{\sigma\mu} \right)
\eeq
\rmrkold{For small $\mu$ one obtains ${s_{\mu} \approx (1/6) \mu \sigma^2}$, 
while in contrast the threshold ${s_{\infty}=\sigma}$ is finite because the distribution 
of the ``activation energies" is bounded.}

\begin{figure}
\includegraphics[height=6cm]{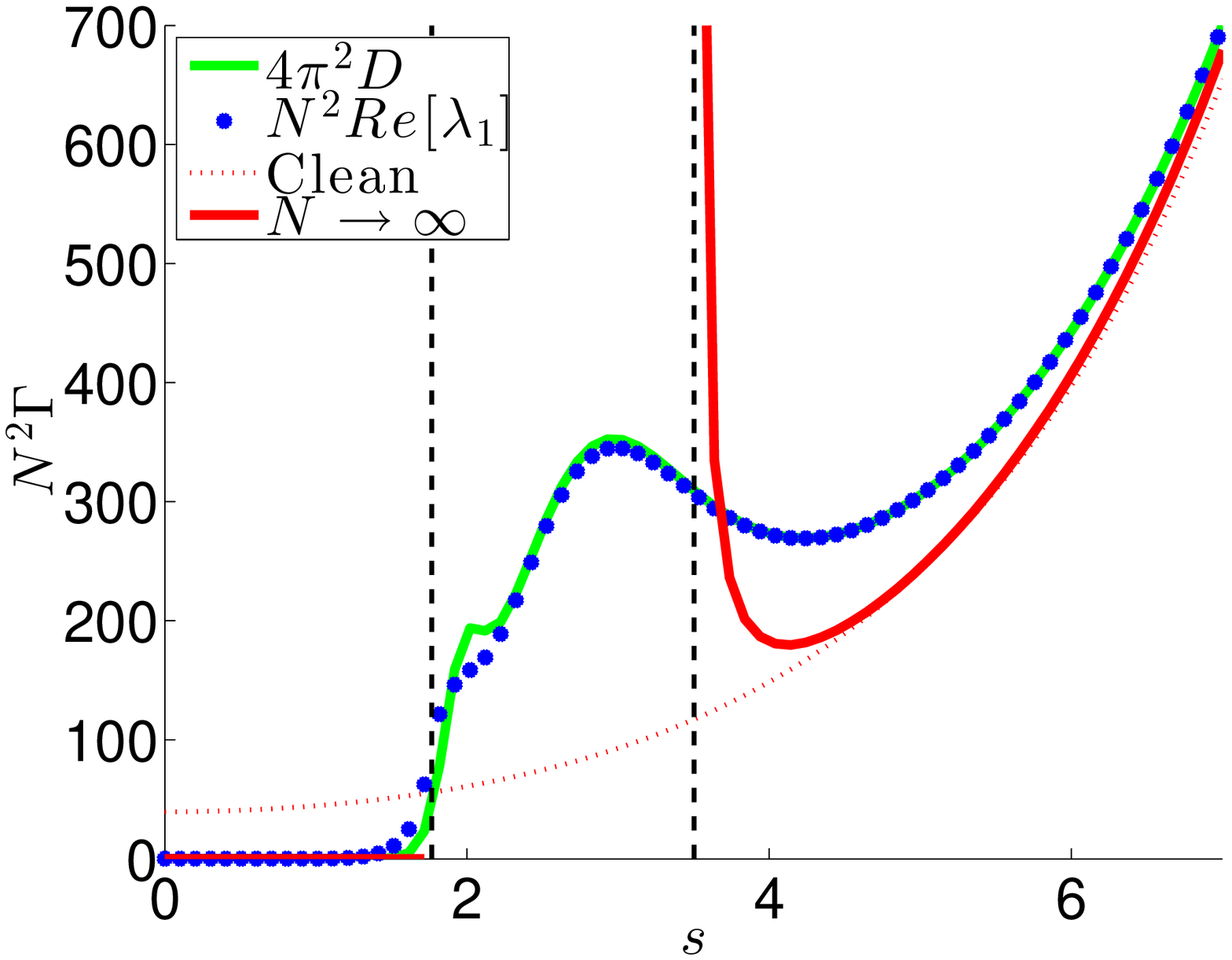} \ \ \ \ 
\includegraphics[height=6cm]{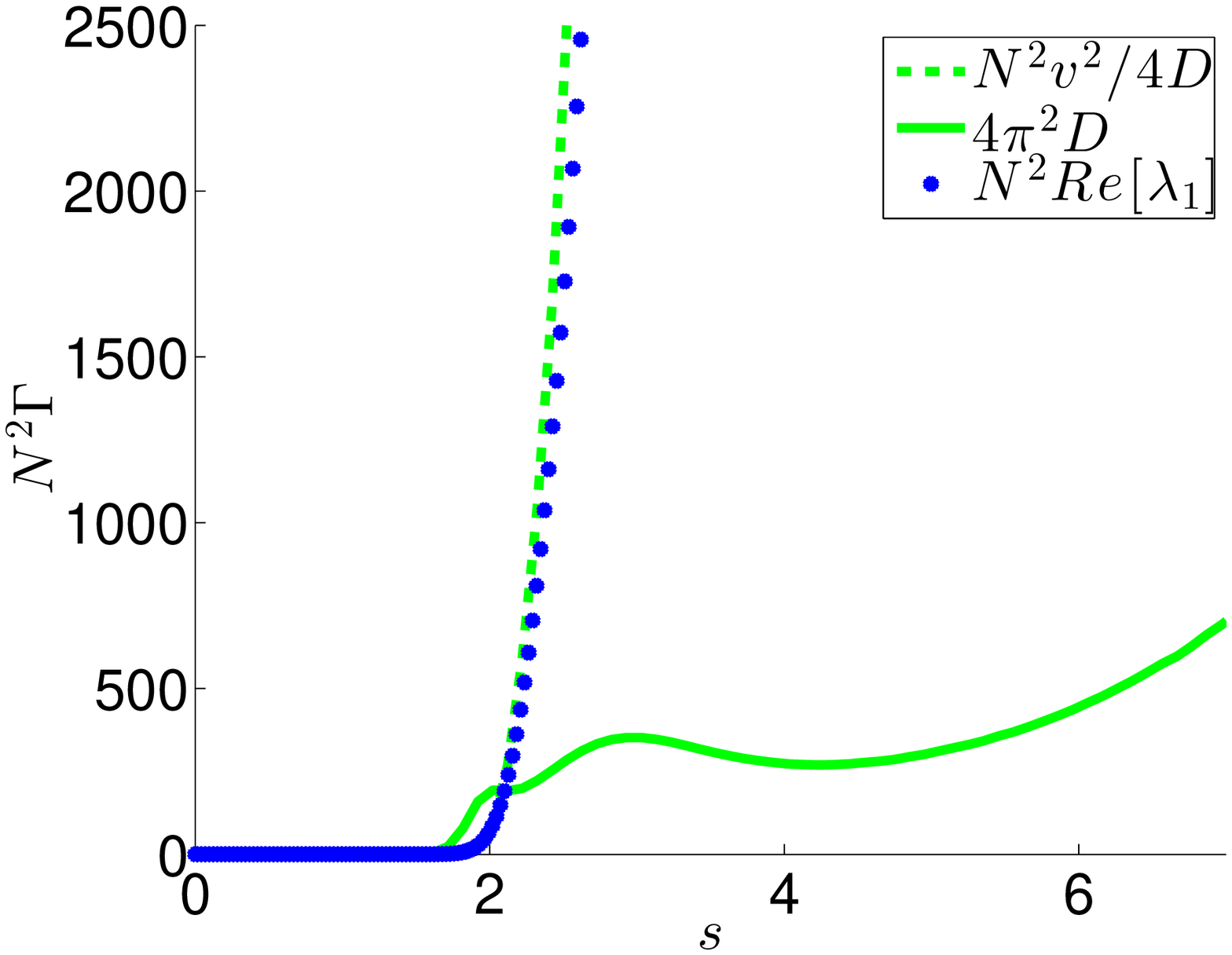}
\caption{
The relaxation rate ${\Gamma = \re[\lambda_1]}$ 
versus the affinity $s$ 
for a sample with ${N=1000}$ sites, 
and disorder strength ${\sigma=5}$. 
\rmrkold{The units of times are chosen such that ${w{=}1}$}.
The upper panel is for a ring, 
while for the lower panel one bond has been disconnected (``box"). 
The blue data points have been obtained 
via numerical diagonalization of the $\bm{W}$ matrix, 
whereas the solid and dashed green lines 
are based on \Eq{e13} and \Eq{e14} 
with numerically calculated~$D$ and~$v$.
\rmrk{The red-dotted-line and the red-thick-solid-line in the upper panel}  
are based on analytical estimates for $D$, 
namely, \Eq{e106} and \Eq{eA1}.   
\rmrkold{The vertical dashed lines are the thresholds $s_{1/2}$ (left) and $s_2$ (right).
The former determines $S_c$ via \Eq{e201}.}}
\label{f1}
\end{figure}

The relaxation spectrum of a \rmrk{finite~$N$ disordered sample (ring or box of length ${L=Na}$)}  
can be found numerically by solving the characteristic equation
\be{16}
\det(z+\bm{W}) \ \ = \ \ 0
\eeq
The relaxation rate is defined as in \Eq{e564}.
For a given realization of disorder we regard $S_{\circlearrowleft}$ 
as a free parameter. Making $S_{\circlearrowleft}$ larger  means that all the $\mathcal{E}_n$ 
are increased by the the same constant. We define the complexity threshold $S_c$  
as the value beyond which the spectrum becomes complex. This means that 
for ${S_{\circlearrowleft} < S_c}$ the relaxation is over-damped like in a box, 
while for ${S_{\circlearrowleft} > S_c}$ the relaxation is under-damped \rmrk{like in a clean ring}.
It has been established \cite{neg} that 
\be{201}
S_c \ \ = \ \ N \ s_{1/2} 
\eeq
In the upper panel of \Fig{f1} we calculate the dependence 
of $\Gamma$ on $S_{\circlearrowleft}$ for a representative 
disordered ring via direct diagonalization of the $\bm{W}$ matrix.
\rmrkold{The results are displayed as blue symbols. 
The complexity threshold \Eq{e201} is indicated by the left 
vertical dashed line.}
In the lower panel of \Fig{f1} we calculate the relaxation rate $\Gamma$  
for a box configuration, i.e. one link of the ring has been disconnected. 
For such configuration the topological aspect is absent and 
therefore the spectrum \rmrk{of the $N$ site sample} is real (${S_c=\infty}$).

We test whether \Eq{e13} and \Eq{e14} can be used in order to predict $\Gamma$. 
For this purpose $v$ and $D$ are independently calculated 
using a standard procedure that is outlined \rmrkold{in Appendix~A of} \cite{nes}. 
\rmrkold{Indeed we observe in \Fig{f1} a nice agreement between 
this prediction (solid and dashed green lines) and 
the previously calculated relaxation rate (blue symbols).}

Having realized that $\Gamma$ of a ring is determined by~$D$ 
\rmrk{via \Eq{e13}}
we would like to understand theoretically the observed non-monotonic 
variation as a function of~$s$. 
In the ${N\to\infty}$ limit the calculation of~$D$ 
can be carried out analytically \cite{odh1}, 
\rmrk{using \Eq{eA1} of} Appendix~\ref{AppA}. 
In this limit ${D=0}$ in the range ${s<s_{1/2}}$ where the spectrum is real; 
then it becomes infinite for ${s_{1/2} < s < s_{2}}$, 
and finite for ${s>s_{2}}$. 
\rmrkold{The result of the calculation in the latter regime 
is represented by the red curve in the upper panel of \Fig{f1}.}
As expected it provides a good estimate 
only for large~$s$ 
\rmrk{where \Eq{eA1} can be approximated by \Eq{e60}}, 
leading to   
\be{52}
\Gamma  \ \ \approx \ \  
\left(\frac{2\pi}{N}\right)^2
\frac{w}{2}\exp\left[ \frac{1}{2}s-\frac{3}{2}s_{1/2}+s_1 \right]
\eeq
Note that this expression roughly coincides 
with the clean ring result \Eq{e13} with \Eq{e106}, 
\rmrkold{see black curve in the upper panel of \Fig{f1}}.

In the range ${s_{1/2} < s < s_{2}}$ the diffusion coefficient 
is large but finite and becomes $N$ dependent. 
In \cite{nes} a heuristic  approach has been attempted 
in order to figure out this $N$ dependence. 
In the present work we would like to adopt a more rigorous 
approach. We shall deduce the $N$ dependence of $\Gamma$ 
analytically from the characteristic equation \Eq{e16}.
We shall also provide an optional derivation for \Eq{e52}.

\begin{figure*}
\includegraphics[height=8cm]{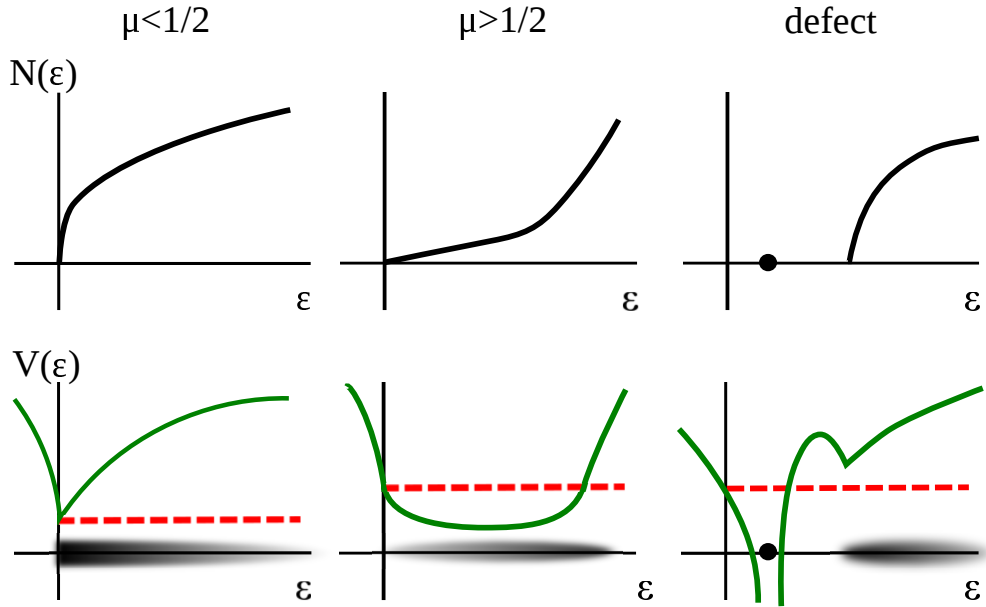}
\caption{Caricature of the electrostatic picture used to determine the transition to complexity.
The panels of the top row display the integrated density of states that comes from~$\varrho(\epsilon)$.
The latter is represented by a cloud along the axes of the lower panels. 
A weak-link \rmrkold{(`defect')} contributes an isolated charge at the vicinity of the origin, 
unlike full disorder \rmrkold{(left panels)} that fills the gap with some finite density. 
The associated envelope of the electrostatic potential is displayed as green lines.  
The dashed red line is $V(0)$. \rmrkold{For ${s<s_{1/2}}$ the spectral density 
has exponent $\mu<1/2$, hence $V'(0)$ is positive, and consequently 
the equation ${V(x)=V(0)}$ has real solutions. 
For  ${s>s_{1/2}}$ the spectral exponent $\mu>1/2$ implies negative $V'(0)$, 
and consequently complex roots appear.}}
\label{figEpp}
\end{figure*}


\section{Extracting $\Gamma$ from the characteristic equation}

With the $\bm{W}$ of the rate equation \Eq{e1} it is possible to
associate a symmetric real matrix $\bm{H}$ as explained in Appendix~\ref{AppB}. 
The latter has real eigenvalues $-\epsilon_k$ with ${k=0,1,2,3,...}$.  
Using the identity \Eq{e99} \rmrkold{of  Appendix~\ref{AppC}}, 
and setting the units of time such that ${w{=}1}$, 
the characteristic equation \Eq{e16} is
\be{202}
\prod_k \left(z-\epsilon_k(s)\right) 
 = (-1)^N 2\left[\cosh\left(\frac{S_{\circlearrowleft}}{2}\right)-1\right]
\eeq
Taking the log of both sides,  \rmrkold{this equation takes the form ${\Psi(z)=\Psi(0)}$.  
The identification of the right hand side as ${\Psi(0)}$ 
is based on the observation that ${z=\lambda_0=0}$ has to be an eigenvalue,  
corresponding to the steady state solution.
It is illuminating to regard $\Psi(z)$ as the complex potential  
in a two dimensional electrostatic problem:}
\be{230}
\rmrkold{\Psi(z) = \sum_k \ln\left( z - \epsilon_k \right)   
\ \equiv \ V(x,y)+iA(x,y) \ \ \ \ } 
\eeq
where ${z=x+iy}$. The constant ${V(x,y)}$ curves correspond to potential contours,
while the constant ${A(x,y)}$ curves corresponds to stream lines. 
The derivative $\Psi'(z)$ corresponds to the field, 
which can be regarded as either an electric or a magnetic field up to a 90deg rotation.    
On the real axis ($x=\epsilon, \ y=0$), the potential is 
\be{25}
V(\epsilon) = \sum_k \ln\left( |\epsilon - \epsilon_k|\right) \equiv \int \ln \left(|\epsilon-\epsilon' \right|) \varrho(\epsilon')d\epsilon' \ \ \ \ 
\eeq
\rmrkold{The spectral density $\varrho(\epsilon)$ of the eigenvalues ${\{\epsilon_k\}}$ 
is further discussed in Appendix~\ref{AppD}. 
Using the language of the electrostatic picture we regard it as a charge distribution.
For full disorder the density for small~$\epsilon$
is characterized by an exponent~$\mu$ 
namely,  ${\varrho(\epsilon)\propto \epsilon^{\mu-1}}$. 
The spectral exponent~$\mu$ is determined via \Eq{e362}. 
An explicit example for the implied dependence 
of $\mu$ on $s$ is provided by inverting \Eq{e15}. 
One observes that $\mu$ becomes infinite   
as $s$ approaches ${s_{\infty} = \sigma}$. 
For  ${s > s_{\infty}}$ a gap is opened. 
In Appendix~\ref{AppE} we provide some insight with regard to the 
implied electrostatic potential $V(\epsilon)$.} 
The bottom line is summarized by \Fig{figEpp}.  
For full disorder, if ${s < s_{1/2}}$ the envelope at the origin has a
\rmrkold{positive} slope hence the equations ${V(x)=V(0)}$ has real solutions, 
and the relaxation spectrum $\{\lambda_k\}$ comes out real.
For ${s > s_{1/2}}$ the envelope at the origin has a \rmrkold{negative slope, 
hence no real solutions at the bottom of the spectrum, 
and the low eigenvalues become complex. 
Accordingly the threshold $S_c$ for full disorder is determined by \Eq{e201}.}  

We would like to estimate the relaxation rate 
in the non-trivial regime ${S_{\circlearrowleft}>S_c}$, 
where the topology of the ring is reflected. 
Given the spectral density $\varrho(x)$, 
the electrostatic potential is 
\be{21}
V(x,y) \ \ = \ \ \frac{1}{2} \int \ln \left[(x-x')^2 + y^2\right] \varrho(x')dx' 
\eeq
Expanding to second order near the origin, we have 
\beq
V(x,y) \ \ \approx \ \ C_0 - C_1 x +\frac{1}{2}C_2y^2 
\eeq
where the coefficients $C_n$ are defined as 
\be{48}
C_n  \ \ =  \ \ \int_0^{\infty} \frac{1}{\epsilon^n}\varrho(\epsilon)d\epsilon
\eeq
Notice that $C_0 = V(0)$ and $C_1 = E(0)$ are the potential 
and the electrostatic field at the origin. 
To determine the real part of the complex gap it is enough to realize 
that the equipotential contour $V{(x,y)=V(0)}$ is approximately 
a parabola near the origin:  
\be{48}
x \ \ = \ \ \frac{1}{2}\frac{C_2}{C_1} y^2 
\eeq
We define as a reference the field-line $A(x,y)=0$ 
that stretches through the origin along the X axis to~$-\infty$.
The first excited eigenvalue is determined by the intersection 
of the $V{(x,y)=V(0)}$ potential contour with the next field line, 
namely with $A(x,y)=2\pi$.
By definition of the stream function~$A(x,y)$, 
which can be regarded as an application of the Cauchy-Riemann theorem,  
it is equivalent to the requirement of having an enclosed flux
\beq
\int_{0}^{\sqrt{2(C_1/C_2)\Gamma}} \left|\vec{E}(x,y)\right| dy \ \ =  \ \ 2\pi
\eeq
The integrand is approximated by ${|\vec{E}(x,y)| \approx C_1}$,
hence we deduce 
\be{49}
\Gamma \ \ \approx \ \ 2\pi^2 \frac{C_2}{C_1^3} 
\eeq
If all the $C$s are proportional to $N$ it follows that ${\Gamma\propto N^{-2}}$ 
as in the case of a clean diffusive ring. This is indeed the case if ${s>s_{2}}$.
But if ${s<s_{2}}$ we have to be careful about the lower cutoff.
From the quantization condition ${\mathcal{N}(\epsilon)=1}$ 
we deduce that ${\epsilon_1 \propto N^{-1/\mu}}$ and get 
\be{374}
\Gamma \propto N^{-\eta}, 
\hspace{5mm}
\eta = 
\left\{\begin{array}{ll}
\frac{1}{\mu}                      & \mbox{for ${s_{1/2} < s < s_{1}}$} \cr
\left(3-\frac{2}{\mu}\right)       & \mbox{for ${s_{1} < s < s_{2}}$} \cr
2                                  & \mbox{for ${s > s_{2}}$} 
\end{array}\right. 
\eeq 
Comparing with \Eq{e13} we realize that consistency requires to assume 
that ${D \propto N^{(2/\mu)-1}}$ for ${s_{1}<s < s_{2}}$, 
and ${D \propto N^{2-(1/\mu)}}$ for ${s_{1/2}<s < s_{1}}$.
The latter result (but not the former) is in agreement 
with the heuristic approach of \cite{nes}.
In the heuristic approach it has been assumed, 
apparently incorrectly, that the disorder-induced 
correlation-length scales like $N$ throughout 
the whole regime ${s<s_{2}}$, and becomes size-independent 
for ${s>s_{2}}$. Apparently the $N$ dependence 
of the disorder-induced correlation-length    
becomes anomalous within the intermediate range ${s_{1} < s < s_{2}}$.

\begin{figure}
\includegraphics[height=6cm]{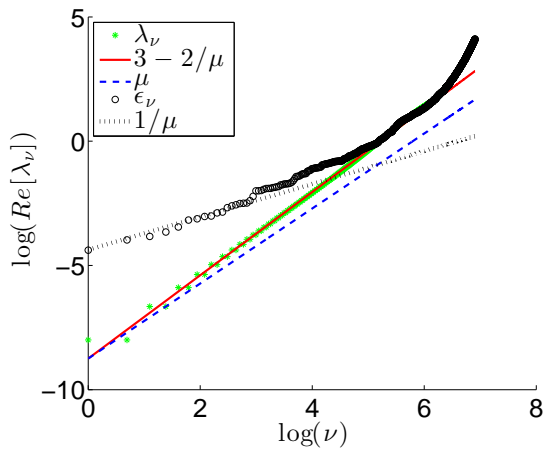}

\caption{
\rmrk{The $\epsilon_{\nu}$ (circles) and the $\re[\lambda_{\nu}]$ (stars) 
versus the index ${\nu=1,2,3...}$ in a natural log-log scale 
for an ${N=1000}$ site disordered ring. The strength of the 
disorder is ${\sigma=5}$, and ${s=3.2015}$ which implies ${\mu = 1.5091}$.
The expected density ${\mathcal{\varrho}(\epsilon) \propto \epsilon^{\mu{-}1}}$ 
at the lower part of the spectrum is confirmed by 
the agreement with ${\epsilon_{\nu}\propto \nu^{1/\mu}}$ (dotted line).  
We compare the numerical result for $\re[\lambda_{\nu}]$
with our prediction ${\eta=3-(2/\mu)}$ (slope of the red solid line), 
and contrast it with the naive heuristic expectation ${\eta=\mu}$ (slope of the blue dashed line).}  
}

\label{f1e}

\end{figure}

\rmrk{
The result \Eq{e374} for $\Gamma$ has an obvious implication on
the spectral density of the relaxation modes.
Clearly $\re[\lambda_{\nu}]$ with ${\nu=0,1,2,3,...}$ 
should be a function of~$\nu/N$,
reflecting that the spectral density is extensive in~$N$.
Accordingly \Eq{e374} can be re-phrased as saying 
that $\re[\lambda_{\nu}] \propto \nu^{\eta}$. 
This result is in general agreement with the heuristic 
argument of \cite{brm2}, but not in the regime ${s_{1} < s < s_{2}}$, 
where it had been argued that ${\eta=\mu}$. 
while our result is ${\eta=3-(2/\mu)}$. 
The maximum difference is for ${\mu\sim1.5}$. 
Our prediction is supported by the numerical example in \Fig{f1e}.
We also note that \Eq{e48} implies that $\im[\lambda_{\nu}] \propto \nu^{\eta/2}$
irrespective of~$\mu$. But a numerical inspection (not displayed) 
shows that the latter approximation  
works well only for the few first eigenvalues. 
}

\section{Sparse disorder}

\rmrkold{So far we have considered ``full disorder" for which one is able in principle 
to determine a coarse-grained diffusion coefficient $D$, that may depend on $L$, 
and from that to extract $\Gamma$ via \Eq{e13}, leading to \Eq{e374}.
But in practice the disorder might be ``sparse" meaning that only a few links 
are defected. The extreme case is having a single ``weak link", meaning a bond or a 
region where the transitions are extremely slow. In such case \Eq{e13} for~$\Gamma$,  
as well as \Eq{e201} for $S_c$ are not physically meaningful. Still we can use the 
electrostatic picture of the previous sections in order to analyze on equal footing 
the secular equation. This will be demonstrated in the subsequent sections.}

\rmrkold{In order to get tangible analytical results we consider a minimal model, 
namely a diffusive ring with a single weak-link region. The length of the diffusive  region is~$L$, and it is characterized by a diffusion coefficient $D_0$, while the length of the defected region is $L_1$, and it is  characterized by a diffusion coefficient $D_1$. We characterize the the weak link region by a ``conductance" parameter $g=(D_1/L_1)/(D_0/L)$, and take the limit ${L_1\to0}$ keeping $g$ constant. We find that the threshold $S_c$ does not depend on~$L$ as in \Eq{e201}, 
but rather reflects~$g$. The characteristic equation implies that $\Gamma \propto 1/L^2$ 
as for a clean ring, but with a prefactor that depends on~$g$. This dependence is illuminating: 
it leads to an interpolation between the ``ring" result \Eq{e13} and the ``box" result \Eq{e14}.}

\section{Ring with weak link}
 
We would like to analyze how the relaxation spectrum is affected 
once a weak-link is introduced into a diffusive ring. 
We use the continuum limit \Eq{e107} for the purpose of deriving 
the characteristic equation.
In a region where $v(x)$ and $D(x)$ are constant 
a free-wave solution ${\rho(x) \propto \eexp{i\tilde{k}x-\lambda t}}$,
has to satisfy the dispersion relation ${\lambda = D\tilde{k}^2 + iv\tilde{k}}$.
It is convenient to use the notation ${s=v/D}$, which would be consistent 
with the discrete-lattice convention if the lattice constant were taken as the unit length.   
Given $\lambda$ we define $k$ that might be either real or pure-imaginary
through the following expression:  
\be{32}
\lambda \ \ \equiv \ \ \left[k^2+ \left(\frac{s}{2}\right)^2 \right]D
\eeq
The complex wavenumbers that correspond to this value 
are ${\tilde{k}_{\pm}=\pm k -i(s/2)}$. 
In each location the actual stationary solution of \Eq{e107}
has to be a superposition 
of clockwise ($k_+$) and anticlockwise ($k_-$) waves
\be{1012}
\rho(x)  
\ &=& \   \Big[ A\eexp{ik x} + B\eexp{-ik x} \Big] \, \eexp{(s/2)x} 
\\ &\equiv& \ \psi^{+}(x) + \psi^{-}(x)
\eeq
We define the state vector 
\beq
\vec{\psi} (x) \equiv \left( \begin{array}{c}
\rho(x) \\
\partial \rho(x)
\end{array}
\right) 
=  
\left(
\begin{array}{cc}
 1 & 1 \\
i\tilde{k}_{+} & i\tilde{k}_{-} \\
\end{array}
\right)
\left( \begin{array}{c}
\psi ^+ (x)\\
\psi ^- (x)
\end{array}
\right)
\eeq
The transfer matrix $M$ that matches the state vector 
at two different locations is defined via the relation
\beq
\vec{\psi}(x_2) \ \ = \ \  M \vec{\psi}(x_1)
\eeq

In a ring with a weak-link there are two segments with different 
diffusion coefficients $D_0$ and $D_1$. 
The continuity of the density $\rho(x)$ and the current ${J=-D(x) \partial \rho(x)+v(x) \rho(x)}$ 
implies that the derivative $\partial \rho$ should have a jump such that across the boundary 
\beq
\left.\left( \begin{array}{c}
\rho \\
\partial \rho
\end{array}
\right)\right|_{1} \ \ = \ \  
\left(
\begin{array}{cc}
 1 & 0 \\
0 & D_0/D_1 \\
\end{array}
\right)
\left.
\left( \begin{array}{c}
\rho \\
\partial \rho
\end{array}
\right) \right|_{0} 
\eeq
We define the matrices 
\beq
U &=& \left(
\begin{array}{cc}
 1 & 1 \\
i\tilde{k}_{+} & i\tilde{k}_{-} \\
\end{array}
\right)
\\
T &=& \left(
\begin{array}{cc}
\eexp{i\tilde{k}_{+} x} & 0 \\
0 & \eexp{i\tilde{k}_{-} x} \\
\end{array}
\right)
\\
R &=& \left(
\begin{array}{cc}
1 & 0 \\
0 & D_0/D_1 \\
\end{array}
\right)
\eeq
For free propagation over a distance~$L$ we have ${M_0=U T_0 U^{-1}}$, 
with $T_0$ that involves a wavenumber~$k$ that is determined by~$D_0$. 
For a weak-link we have ${M_1=R^{-1}UT_1 U^{-1}R}$, 
where $T_1$ describes the free propagation in the $D_1$ region
\rmrkold{that has some length~$L_1$. 
It is convenient to define the effective length of 
the weak link as $\ell = (D_0/D_1)L_1$.} 
The only non-trivial way to take the limit 
of zero thickness weak-link \rmrkold{(${L_1\rightarrow 0}$)} 
is to adjust ${D_1 \rightarrow 0}$ such that $\ell$ is kept constant.
This leads to the following result:
\beq
M_1  \ \ = \ \ R^{-1}UT_1 U^{-1}R \ \ = \ \
\left( \begin{array}{cc}
1 & \ell  \\
0 & 1
\end{array}
\right)
\eeq
The characteristic equation is 
\be{461}
\det\Big[1- M_1 M_0 \Big] \ \ = \ \ 0
\eeq
leading to 
\be{1071}
\cos(q) - \frac{1}{2g} \left[q^2 + \left(\frac{S_{\circlearrowleft}}{2}\right)^2\right]\frac{\sin(q)}{q}  =  \cosh\left(\frac{S_{\circlearrowleft}}{2}\right)
\eeq
where we have defined
\beq
g \ \equiv \ \frac{L}{\ell} =  \frac{D_1/L_1}{D_0/L}
\eeq
along with $q=kL$ and ${S_{\circlearrowleft}=sL}$.

\begin{figure}
\includegraphics[height=6cm]{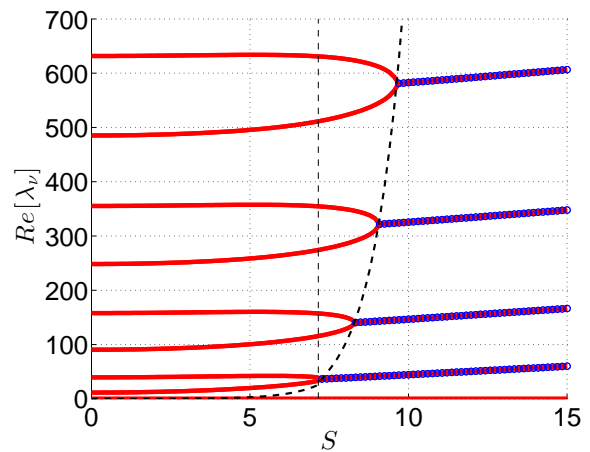}

\caption{
The lower eignevalues for a ring with a weak link \rmrk{versus~$S \equiv S_{\circlearrowleft}$}.
The units of length and time are such that ${D=L=1}$ and we set ${g=0.2}$.
For large enough $S$ the eigenvalues become complex and the real parts coalesce (indicated by blue circles).
The threshold is indicated by the dashed curve that has been 
deduced from the envelope of the characteristic equation.   
The dashed vertical lines indicates $S_c$ of \Eq{e462}.
}

\label{f1s}

\end{figure}

\begin{figure}

\includegraphics[height=6cm]{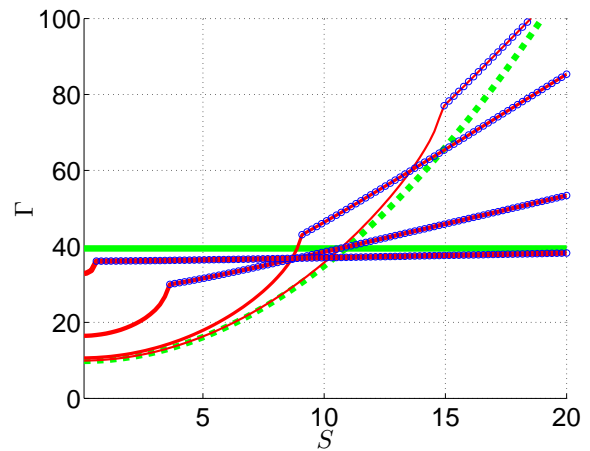}

\caption{
\rmrkold{The relaxation rate $\Gamma$ for the ring of \Fig{f1s} \rmrk{versus~$S \equiv S_{\circlearrowleft}$}.
The horizontal solid green line is for a clean ring (${g=\infty}$), 
while the dashed green line is for a disconnected ring (${g=0}$).
The other lines are for ${g=10,1,0.1,0.01}$. 
To the right of each knee the first eigenvalue ($\lambda_1$) becomes complex, 
indicated by the blue circles. This figure should be  
contrasted with \Fig{f1} -- the significant difference 
is the non-monotonic $s$ dependence there 
(one should not be overwhelmed by the different $s$ dependence 
in the clean-ring limit, see text).}
}

\label{f1c}

\end{figure}

\rmrkold{In \Fig{f1s} we find the dependence of the lowest eigenvalues on $S_{\circlearrowleft}$  
via numerical solution of \Eq{e1071} and using \Eq{e32}.
The units of length and time are such that ${D = L = 1}$. 
From the first eigenvalue we get $\Gamma$ as defined in \Eq{e564}. 
It is implied by the re-scaling of the variables 
in the characteristic equation that $\Gamma \propto 1/L^2$ 
as for a clean ring. 
In \Fig{f1c} we illustrate the dependence of $\Gamma$ on $S_{\circlearrowleft}$ and on $g$.   
The observed $S_{\circlearrowleft}$~dependence is monotonic, unlike that of \Fig{f1}.
In the clean-ring limit there is no $S_{\circlearrowleft}$ dependence 
because we are considering the continuum-limit, 
setting ${D=1}$ irrespective of~$s$, 
while in \Fig{f1} the diffusion coefficient was given by \Eq{e106}. 
Irrespective of this presentation issue, 
as $g$ decreases the \rmrk{drift-determined} $s$~dependence 
is approached, in consistency with \Eq{e14}. 
Thus we have a nice interpolation between the ``ring" and ``box" 
expressions for~$\Gamma$.}

In order to determine the threshold $S_c$ for the appearance of complex eigenvalues   
we take a closer look at \Eq{e1071}. 
The left hand side is an oscillating function within an envelope
\beq
A(q) \ \ = \ \ \sqrt{1+\frac{1}{g^2}\left( \frac{q^2+(S_{\circlearrowleft}/2)^2}{2q}\right)^2}
\eeq
This envelope has a minimum at $q=S_{\circlearrowleft}/2$. 
Accordingly if ${A(S_{\circlearrowleft}/2) < \cosh(S_{\circlearrowleft}/2)}$
complex eigenvalues appear, and we can deduce the  
threshold $S_c$ from the equation 
\be{462}
\sqrt{1+\left(\frac{S_{\circlearrowleft}}{2g}\right)^2} \ \ = \ \  \cosh \left(\frac{S_{\circlearrowleft}}{2}\right)
\eeq
To get an explicit expression we solve the 
approximated equation ${S_{\circlearrowleft}/(2g) = \cosh(S_{\circlearrowleft}/2)}$
and deduce a solution in terms of the Lambert function,
\be{23}
S_c \ \ = \ \ -2 \mathbb{W}(-g/2)
\eeq
This is valid provided $S_{\circlearrowleft} \gg g$, which is self-justified for small~$g$.   
We can use the same procedure in order to determine the 
complexity threshold for a given eigenvalue $\lambda$ in \Fig{f1s}. 
Recall that the corresponding $q$ is $q^2={L^2 \lambda/D_0-S_{\circlearrowleft}^2/4}$.
Solving the quadratic equation ${A(q)=\cosh(S_{\circlearrowleft}/2)}$ 
we find the~$q$ beyond which the spectrum becomes real again.
It terms of $\lambda$ the explicit expression is 
\beq
\lambda_c =
\frac{2D_0}{L^2} g^2 \sinh^2 \left(\frac{S_{\circlearrowleft}}{2}\right) \left[ 
1+\sqrt{1{-}\left(\frac{S_{\circlearrowleft}}{2g\sinh\frac{S_{\circlearrowleft}}{2}}\right)^2}
\right] \ \ \ \ \
\eeq
This boundary is indicated by a dashed black line in \Fig{f1s}).

\begin{figure}
\includegraphics[height=6cm]{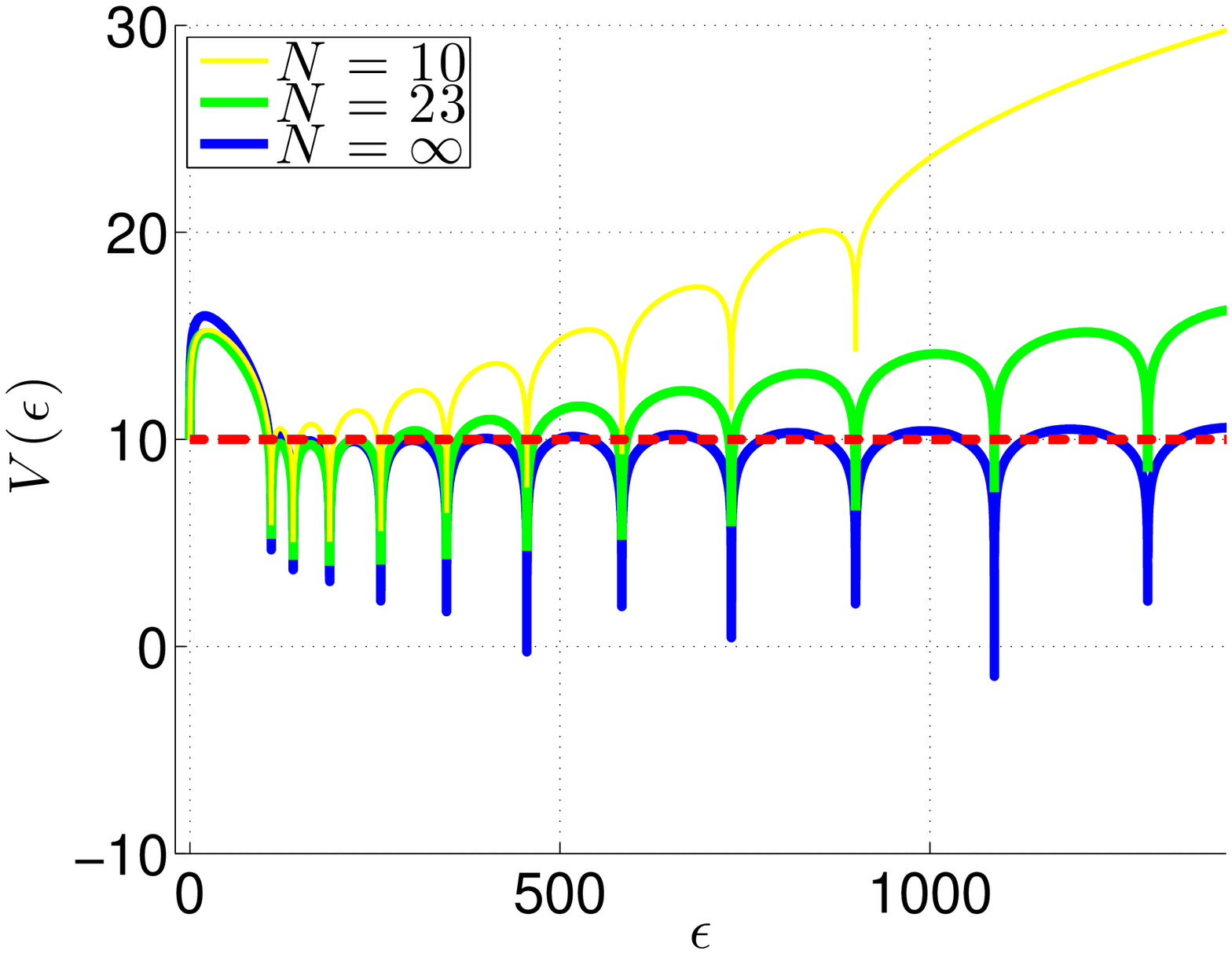} \ \ \ \ \
\includegraphics[height=6cm]{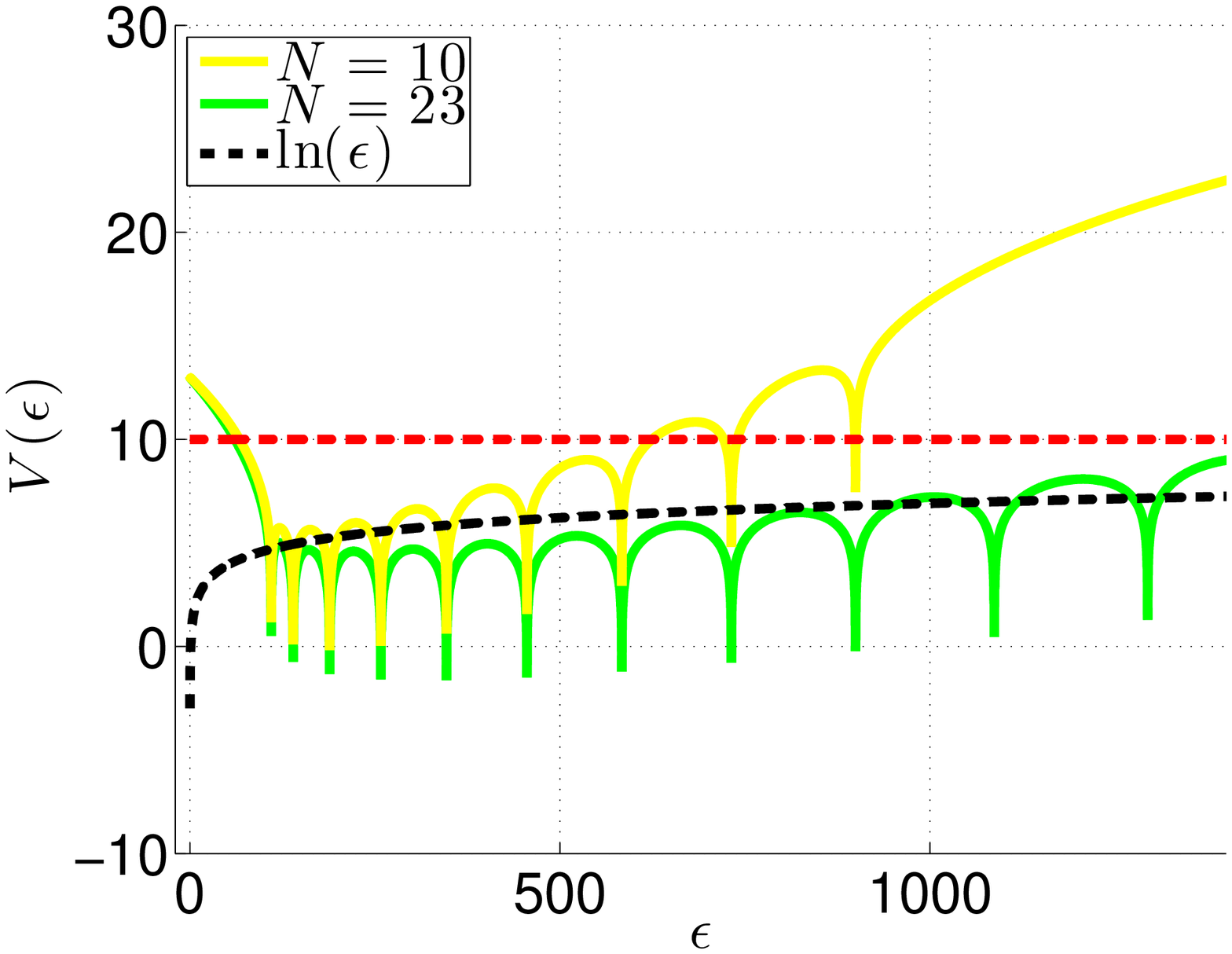}
\caption{
Electrostatic reconstruction of the characteristic equation of a continuous 
ring with weak-link with ${D=L=1}$ and $g=10^{-3}$ and $S_{\circlearrowleft}=20$. 
The blue line is the electrostatic potential of a continuous ring with a defect. 
The dashed red line is $V(0)$. The yellow and green lines 
are reconstructions using a finite number of (numerically obtained) charges.
By increasing the number of charges that are included in the reconstruction, 
it is clear that the deviation from the blue line is due to finite size truncation. 
In the lower panel we display the contribution of the impurity-level charge (dashed black line) 
and the quasi-continuum charges (the other lines) to the reconstructed potential.
}
\label{figReconstruction}
\end{figure}

\section{Reconstruction of the continuum limit}

By reverse engineering, requiring consistency between \Eq{e1071} and \Eq{e202}, 
we deduce that the electrostatic potential that is associated
with the characteristic equation for a ring with a weak link is
\beq
V(\epsilon) = \ln\left\{ 2(\cos(q){-}1) - \frac{1}{g} \left[q^2 {+} \left(\frac{S_{\circlearrowleft}}{2}\right)^2\right]\frac{\sin(q)}{q} \right\} 
\hspace*{8mm} 
\eeq  
This potential is plotted in \Fig{figReconstruction}, and labeled as ``${N{=}\infty}$".   
We would like to reconstruct this potential by means of \Eq{e25}.
For this purpose we have to find the real eignevalues 
of the associated $\bm{H}$, see \Eq{eB6}. 
Formally the equation ${\det(z+\bm{H})=0}$ is obtained by setting $S_{\circlearrowleft}{=}0$ 
in the right hand side (RHS) of \Eq{e1071}, leading to    
\be{1021}
\cos(q) - \frac{1}{2g} \left[q^2 + \left(\frac{S_{\circlearrowleft}}{2}\right)^2\right]\frac{\sin(q)}{q} \ \ = \ \ 1
\eeq
From \Eq{e32} it follows that ${\epsilon_k = \left[q_k^2 +S_{\circlearrowleft}^2/4\right]D_0/L^2}$, 
where $q_k$ are the roots of the above equation.  
Using these ``charges" we compute $V(\epsilon)$ via \Eq{e25}
and plot the result in the upper panel of \Fig{figReconstruction}.
Some truncation is required, so we repeat the attempted reconstruction
with $N=10$ and $N=23$ roots. We observe that the result 
converges to the ${N{=}\infty}$ limit.
The residual systematic error as $\epsilon$ becomes larger 
is due to finite truncation of the number of roots used in the reconstruction.

The characteristic, equation \Eq{e1071} parallels the discrete version \Eq{e202}.
One should be aware that the spectral density contains 
an ``impurity" charge $\epsilon_0$ as illustrated in the third panel of \Fig{figEpp}.
It is easy to explain the appearance of this exceptional charge 
using the discrete-lattice language. In the absence of a weak link 
the diagonal elements of the $\bm{W}$ matrix are ${-\gamma}$ 
where ${\gamma = w\eexp{s/2}+w\eexp{-s/2}=2w\cosh{s/2}}$.   
The spectrum of the associated $\bm{H}$ matrix forms a band, 
such that the lower edge of $\varrho(\epsilon)$ is 
\beq
\epsilon_{\mbox{floor}} \ \ = \ \ \gamma-2w \ \ = \ \ 2w \left[\cosh(s/2)-1\right]
\eeq
If we introduce a weak-link ${w_{0}\ll w}$ at the ${(0,1)}$ bond, 
we get one exceptional diagonal element ${\gamma_0}$.
Consequently, for small enough $w_{0}$, there is an 
out-of-band impurity level that does not mix with the band:
\beq
\epsilon_0 \ \ \approx \ \ \gamma_0 \ \ =  w_0\eexp{s/2}+w\eexp{-s/2}
\eeq
In the lower panel of \Fig{figReconstruction} 
we separate the contribution of the impurity level 
from the contribution off all the other band-levels.

\section{Discussion}

We have outlined a physically appealing procedure to extract the relaxation rate 
of a stochastic spreading process in a closed ring, bridging between the discrete model  
and its continuum limit, and treating on equal footing full and sparse disorder.
By sparse disorder we mean several weak-links. For presentation purposes we have provided a
full analysis for a ring with a single defect, but the generalization to 
several weak links is merely a technical issue. 

Our approach has been inspired by previous works regrding non-Hermitian Hamiltonians \cite{Hatano1,Hatano2,Shnerb1}, 
and follows our previous publication \cite{neg} regarding the determination of the complexity threshold. 
In the present work the emphasis was on the determination of the relaxation rate $\Gamma$ 
in the ``complex" regime where the topological aspect manifests itself.
Generally speaking in this regime $\Gamma$ may exhibit anomalous dependence on the length of the sample. 

\ \\

\noindent
{\bf Acknowledgements.-- }
We thank Oleg Krichevsky (BGU) for a helpful discussion.
This research has been supported by  by the Israel Science Foundation
(grant No. 29/11).

\newpage
\appendix

\section{Expressions for $v$ and $D$ in the presence of disorder }
\label{AppA}
In the presence of disorder, the forward and backward rates are random numbers.  
Here we summarize known analytical expressions for $v$ and $D$ based on \cite{odh1}, 
and notations as in \cite{nes,neg}. 
Taking the infinite chain limit, 
and using units such that the lattice spacing is ${a=1}$, 
the expression for the drift velocity is  
\beq
v \ \ = \ \ \frac{1- \left\langle  \frac{\overleftarrow{w}}{\overrightarrow{w}}\right\rangle }{\left\langle  \frac{1}{\overrightarrow{w}}\right\rangle} 
\eeq
We notice that a non-percolating resistor-network-disorder will diminish the drift velocity as expected due to the denominator. 
Irrespective of that the result above is valid only in the ``sliding regime" where ${v>0}$. 
Looking at the numerator one observes that the implied condition for that is ${s>s_1}$. 
As for the diffusion, it becomes finite for ${s>s_2}$, and the explicit expression is  
\be{A1}
D \ = &&  \frac{1-\left\langle  \frac{\overleftarrow{w}}{\overrightarrow{w}}\right\rangle^2}{1- \left\langle  \left(\frac{\overleftarrow{w}}{\overrightarrow{w}}\right)^2\right\rangle}
\left\langle  \frac{1}{\overrightarrow{w}}\right\rangle^{-3} 
\\ \nonumber &\times& 
\left[ 
\left\langle  \frac{1}{\overrightarrow{w}}\right\rangle  \left\langle  \frac{\overleftarrow{w}}{\overrightarrow{w}^2}\right\rangle+\frac{1}{2} \left\langle  \frac{1}{\overrightarrow{w}^2}\right\rangle
\left( 1 - \left\langle  \frac{\overleftarrow{w}}{\overrightarrow{w}}\right\rangle \right)
\right] 
\eeq
For large bias a practical approximation is 
\be{60}
D \ \ \approx \ \
\frac{1}{2} \left\langle \frac{1}{\ora w}\right\rangle ^{-3}   \left\langle \frac{1}{\ora{w}^2}\right\rangle 
\eeq
Considering a ring with random rates ${w\eexp{\pm \mathcal{E}_n/2}}$, 
the dependence of all the various expectation values 
on the affinity~$s$ is expressible in terms of the 
parameters $w$ and $s_{\mu}$. For example 
\beq
v \ \ = \ \ \eexp{\frac{1}{2}(s_1-s_{1/2})} \left[ 2 \sinh\left(\frac{s-s_1}{2}\right) \right] w
\eeq

\section{The associated $H$ matrix}
 \label{AppB}
Our model is described by a conservative matrix $\bm{W}$ that describes hopping between sites. 
In the Chain configuration the site index $n$ runs from $-\infty$ to $\infty$, 
while in the Ring configuration it is defined modulo~$N$. 
In the latter case we characterize the stochastic field by a potential $U(n)$ and 
by an affinity $S_{\circlearrowleft}$, such that 
\beq
\mathcal{E}_n \ \ = \ \ U(n)-U(n{-}1) \, + \, \frac{S_{\circlearrowleft}}{N}
\eeq
Then we associate with $\bm{W}$ a similar matrix~$\tilde{\bm{W}}$ 
and a real symmetric matrix~${\bm{H}}$ as follows: 
\beq \nonumber 
{\bm{W}} &=&   
\mbox{diagonal}\Big\{-\gamma_{n}({s}) \Big\} 
+\mbox{offdiagonal}\Big\{  w_{n}\eexp{\pm \frac{{\mathcal{E}_n}}{2}}  \Big\} 
\\ \nonumber 
\tilde{\bm{W}} &=& 
\mbox{diagonal}\Big\{-\gamma_{n}({s})\Big\} 
+\mbox{offdiagonal}\Big\{  w_{n}\eexp{\pm \frac{{S_{\circlearrowleft}}}{2N}}  \Big\} 
\\ \nonumber 
{\bm{H}} &=& 
\mbox{diagonal}\Big\{-\gamma_{n}({s})\Big\} 
+\mbox{offdiagonal}\Big\{  w_{n}\Big\} 
\eeq
such that 
\beq
\tilde{\bm{W}} \ \ = \ \ \eexp{\bm{U}/2} \bm{W} \eexp{-\bm{U}/2}
\eeq
where ${\bm{U}=\mbox{diag}\{ U(n) \} }$ is a diagonal matrix.
The relation between ${\bm{W}}$ and $\tilde{\bm{W}}$ can be regarded 
as a gauge transformation, and $S_{\circlearrowleft}$ can be regarded 
as an imaginary Aharonov-Bohm flux. The hermitian matrix ${\bm{H}}$ can be regarded 
as the Hamiltonian of a particle in a ring in the absence of a magnetic flux.  
The $\bm{W}$ of a clean ring \Eq{e3} and its associated~$\bm{H}$ are  
\beq
{\bm W} \ &=& \ 2w \left[\cos\left(\bm{P}+i\frac{s}{2}\right) - \cosh\left(\frac{s}{2}\right) \right]
\\\label{eB4}
{\bm H} \ &=& \ 2w \left[\cos\left(\bm{P} \right) - \cosh\left(\frac{s}{2}\right) \right]
\eeq
while in the continuum limit \Eq{e107} implies that 
\beq
{\bm W} \ &=& \ -D \bm{P}^2 \ + \ iv\bm{P} 
\\ \label{eB6}
{\bm H} \ &=& \ -D \left[ \bm{P}^2 + \left( \frac{v}{2D} \right)^2 \right]
\eeq
In the absence of disorder the eignevalues are obtained
by the simple substitution ${\bm{P} \mapsto (2\pi/L)\nu}$, 
where $\nu$ is an integer.

\section{The characteristic equation}
\label{AppC}

Consider the tridiagonal matrix  
\beq
\bm{A} \ \ = \ \ 
\left(
\amatrix{
a_0  & b_1 & 0   & ... & c_0 \cr
c_1  & a_1 & b_2 & ... &  0   \cr
0    & c_2 & a_2 & ... &  0   \cr
...  & ... & ... & ... & ...  \cr
b_0  &  0  &  0  & ... &  0   \cr
} \right)
\eeq
and associated set of transfer matrices
\beq
T_n &=& \left(
\amatrix{
a_n & -b_n c_n \cr
1 & 0 
}
\right)
\eeq
Our modified indexing scheme of the elements, 
allows a simpler presentation of the formula 
for the determinant that appears in \cite{det1}: 
\beq \nonumber
\det[\bm{A}] =  
\mathrm{trace} \left[\prod_{n=1}^N T_n \right] 
- (-1)^N \left[ \prod_{n=1}^N b_n + \prod_{n=1}^N c_n  \right]  
\eeq
From here follows 
\be{99}
&& \det (z+\bm{W})  
\ \ = \ \  \det (z+\tilde{\bm{W}}) 
\\ \nonumber &&
\ = \det (z+{\bm{H}}) -  2\left[\cosh\left(\frac{{S_{\circlearrowleft}}}{2}\right)-1\right](-w)^N
\eeq
Hence the characteristic equation is \Eq{e202}.

\section{The spectral density $\varrho(\epsilon)$}
\label{AppD}

Consider a ring where the transition rates between neighboring sites 
are random variables ${w\eexp{\pm \mathcal{E}_n/2 }}$.    
The equation that describes the relaxation in such a ring 
in the continuum limit is \Eq{e107} with ``white disorder". 
Namely $v(x)$ has Gaussian statistics with ${\braket{v(x)v(x')}=\nu_{\sigma} \delta(x{-}x')}$   
where ${\nu_{\sigma}=w^2a^3\mbox{Var}(\mathcal{E})}$.  
Assuming $D(x)=D_0$, and adding to the disorder an average value~$v_0$, 
one observes that the diffusion equation is characterized by 
a single dimensionless parameter. It is customary to define 
\rmrkold{in consistency with \Eq{e15}}  
\be{D1}
\mu \ \ \equiv \ \ \frac{2D_0}{\nu_{\sigma}} v_0 \ \ = \ \ \frac{2s}{\mbox{Var}(\mathcal{E})} 
\eeq   
This parameter equals $v_0$ if we use the common re-scaling 
of units such that ${2D_0 = \nu_{\sigma} =1 }$. 
Then the units of time and of length are 
\beq
\ [T] \ &=& \ \frac{8D_0^3}{\nu_{\sigma}^2} \ = \ \left[\frac{8}{\mbox{Var}(\mathcal{E})^2}\right]w^{-1} 
\\ 
\ [L] \ &=& \ \frac{4D_0^2}{\nu_{\sigma}} \ = \ \left[\frac{4}{\mbox{Var}(\mathcal{E})}\right]a   
\eeq

In the absence of disorder, by inspection of \Eq{eB6}, 
the spectral density $\varrho(\epsilon)$ is like that 
of a ``free particle" but shifted upwards such that 
the band floor is ${\epsilon_0 = (1/4) v^2/D}$. 
In the presence of Gaussian disorder the gap ${[0,\epsilon_0]}$ 
is filled. In scaled units the integrated density 
of states is \cite{odh3}:
\beq
\mathcal{N}(\epsilon) \ \ = \ \ \frac{1}{\pi^2}\frac{L}{J_{\mu}^2(\sqrt{2\epsilon})+Y_{\mu}^2(\sqrt{2\epsilon})}
\eeq
where $J_{\mu}$ and $Y_{\mu}$ are Bessel functions of the first and second kind. 
For any ${\mu}$ the \rmrkold{large~$\epsilon$ asymptotics}  
gives ${\mathcal{N}(\epsilon) \approx (1/\pi)\sqrt{2\epsilon}}$ 
in agreement with the free particle result.     
In the other extreme, for small $\epsilon$ we get ${\mathcal{N}(\epsilon) \propto \epsilon^{\mu}}$.
It is also not difficult to verify that the 
clean ring spectrum (with its gap) is recovered in the ${\sigma\mapsto0}$ limit.

We have verified that for box-distributed $\mathcal{E}_n$ the 
approximation ${\mathcal{\varrho}(\epsilon) \propto \epsilon^{\mu-1}}$  
holds at the vicinity of the band floor. 
In contrast with a Gaussian distribution $\mu$ becomes infinite   
as $s$ approaches ${s_{\infty} = \sigma}$, see \Eq{e15}. 
For  ${s > s_{\infty}}$ a gap is opened.

\section{Step by step electrostatics}
\label{AppE}

The eigenvalues $\epsilon_n$ of $\bm{H}$ can be regarded as the locations 
of charges in a 2D electrostatic problem. 
We would like to gain some intuition for the associated potential along the real axis. 
For a point charge at $\epsilon_1$ we have ${V(\epsilon) = \ln |\epsilon-\epsilon_1|}$. 
For a uniform charge distribution within $\epsilon \in [a,b]$ we get  
\beq
&& V(\epsilon) \ \ = \ \ \frac{1}{b-a}\int_a^b \ln |\epsilon-\epsilon'| \ d\epsilon' 
\\ \nonumber 
&& \ = \frac{1}{b{-}a}\left[(\epsilon{-}a)\ln|\epsilon{-}a| - (\epsilon{-}b)\ln |\epsilon{-}b| + (a{-}b) \right] 
\eeq
which has a minimum at $\epsilon=(a+b)/2$ and resembles a ``soft well" potential.
In order to have a flat floor the density has to be larger at the edges.
This is the case for a charge density that corresponds to the 
spectrum of a clean ring. The locations of the charges are  
\beq
\epsilon_n = 2\left[ \cosh \left(\frac{s}{2}\right) - \cos\left(\frac{2\pi}{N}n\right) \right] \ \equiv \ \epsilon(k_n)
\eeq
and the potential along the real axis is  
\beq
V(\epsilon) \ \ = \ \ \frac{N}{2\pi} \int_0^{2\pi} \ln \left|\epsilon - \epsilon(k) \right| \ dk 
\eeq
For $\epsilon$ within the band, the integrand can be written 
as $\ln|2(\cos(k_0)-\cos(k))|$, and accordingly the potential vanishes,  
reflecting an infinite localization length.

In the continuum limit the charge density in the case 
of a clean ring behaves as ${\varrho(\epsilon)\propto \epsilon^{\mu-1}}$   
with ${\mu=1/2}$ and leads to a flat floor. 
For general $\mu$ one can show \cite{neg} that
\be{19}
V'(\epsilon) \ \ \propto \ \ \pi \mu \cot(\pi \mu) \ \epsilon^{\mu-1}
\eeq
such that the sign of $V'(\epsilon)$ is positive for ${\mu<1/2}$, 
and negative for ${\mu>1/2}$. See \Fig{figEpp} for an illustration.  
We also illustrate there what happens if we have a clean ring that is perturbed 
by a defect that contributes a charge in the gap.

For $s>s_{\infty}$ we have ${\mu=\infty}$, meaning that 
a gap is opened.  If $s$ is sufficiently large 
the eigenstates of ${\bm H}$ are ``trivially localized",  
so the eigenvalues are simply
\beq
\epsilon_n \ \ = \ \ \exp[(s+\varsigma_n)/2] 
\eeq
where $\varsigma_n \in [-\sigma,\sigma]$ is uniformly distributed.
Accordingly the charge density is ${\varrho(\epsilon) = N/\sigma\epsilon}$
within an interval ${\epsilon \in [a,b]}$, where ${a=\exp[(s-\sigma)/2]}$ and ${b=\exp[(s+\sigma)/2]}$,
leading to
\beq \nonumber
V(\epsilon) \ = \  && \frac{N}{\sigma}\left[
\ln(|\epsilon-a|) \ln\left(\frac{\epsilon}{a}\right) 
-
\ln(|\epsilon-b|) \ln\left(\frac{\epsilon}{b}\right)  \right.
\\
&& 
\left.
+ \mbox{Li}_2\left( 1-\frac{a}{\epsilon}\right)
+ \mbox{Li}_2\left( 1-\frac{b}{\epsilon}\right) \right] 
\eeq
We would like to calculate the decay rate as described by \Eq{e49}.
To carry out the calculation it is easier to integrate with respect to $\varsigma$.
Expanding \Eq{e21} in the vicinity of the origin we get the coefficients 
\beq \nonumber 
C_1  \ \ &=& \ \ \frac{N}{2\sigma} \int_{-\sigma}^{\sigma} e^{-(s+\varsigma)/2}d\varsigma \\
&=& \ \frac{2N}{\sigma} \sinh \left( \frac{\sigma}{2} \right) e^{-s/2} \ = \ Ne^{(s_{1/2}-s)/2} 
\\ \nonumber 
C_2 \ \ &=& \ \ \frac{N}{2\sigma} \int_{-\sigma}^{\sigma} e^{-(s+\varsigma)}d\varsigma  \\
&=& \ \frac{N}{\sigma}\sinh(\sigma)e^{-s}  \ = \ Ne^{s_1-s}
\eeq  
Substitution of $C_1$ and $C_2$ into \Eq{e49} leads to a result that agrees with \Eq{e52}.


\onecolumngrid

\begin{thebibliography}{19}%
\makeatletter
\providecommand \@ifxundefined [1]{%
 \@ifx{#1\undefined}
}%
\providecommand \@ifnum [1]{%
 \ifnum #1\expandafter \@firstoftwo
 \else \expandafter \@secondoftwo
 \fi
}%
\providecommand \@ifx [1]{%
 \ifx #1\expandafter \@firstoftwo
 \else \expandafter \@secondoftwo
 \fi
}%
\providecommand \natexlab [1]{#1}%
\providecommand \enquote  [1]{``#1''}%
\providecommand \bibnamefont  [1]{#1}%
\providecommand \bibfnamefont [1]{#1}%
\providecommand \citenamefont [1]{#1}%
\providecommand \href@noop [0]{\@secondoftwo}%
\providecommand \href [0]{\begingroup \@sanitize@url \@href}%
\providecommand \@href[1]{\@@startlink{#1}\@@href}%
\providecommand \@@href[1]{\endgroup#1\@@endlink}%
\providecommand \@sanitize@url [0]{\catcode `\\12\catcode `\$12\catcode
  `\&12\catcode `\#12\catcode `\^12\catcode `\_12\catcode `\%12\relax}%
\providecommand \@@startlink[1]{}%
\providecommand \@@endlink[0]{}%
\providecommand \url  [0]{\begingroup\@sanitize@url \@url }%
\providecommand \@url [1]{\endgroup\@href {#1}{\urlprefix }}%
\providecommand \urlprefix  [0]{URL }%
\providecommand \Eprint [0]{\href }%
\providecommand \doibase [0]{http://dx.doi.org/}%
\providecommand \selectlanguage [0]{\@gobble}%
\providecommand \bibinfo  [0]{\@secondoftwo}%
\providecommand \bibfield  [0]{\@secondoftwo}%
\providecommand \translation [1]{[#1]}%
\providecommand \BibitemOpen [0]{}%
\providecommand \bibitemStop [0]{}%
\providecommand \bibitemNoStop [0]{.\EOS\space}%
\providecommand \EOS [0]{\spacefactor3000\relax}%
\providecommand \BibitemShut  [1]{\csname bibitem#1\endcsname}%
\let\auto@bib@innerbib\@empty
\bibitem [{\citenamefont {Sinai}(1983)}]{Sinai}%
  \BibitemOpen
  \bibfield  {author} {\bibinfo {author} {\bibfnamefont {Y.~G.}\ \bibnamefont
  {Sinai}},\ }\href {\doibase 10.1137/1127028} {\bibfield  {journal} {\bibinfo
  {journal} {Theor. Probab. Appl.}\ }\textbf {\bibinfo {volume} {27}},\
  \bibinfo {pages} {256} (\bibinfo {year} {1983})},\ \Eprint
  {http://arxiv.org/abs/http://dx.doi.org/10.1137/1127028}
  {http://dx.doi.org/10.1137/1127028} \BibitemShut {NoStop}%
\bibitem [{\citenamefont {Derrida}(1983)}]{odh1}%
  \BibitemOpen
  \bibfield  {author} {\bibinfo {author} {\bibfnamefont {B.}~\bibnamefont
  {Derrida}},\ }\href {\doibase 10.1007/BF01019492} {\bibfield  {journal}
  {\bibinfo  {journal} {J. Stat. Phys.}\ }\textbf {\bibinfo {volume} {31}},\
  \bibinfo {pages} {433} (\bibinfo {year} {1983})}\BibitemShut {NoStop}%
\bibitem [{\citenamefont {Bouchaud}\ \emph {et~al.}(1990)\citenamefont
  {Bouchaud}, \citenamefont {Comtet}, \citenamefont {Georges},\ and\
  \citenamefont {Doussal}}]{odh3}%
  \BibitemOpen
  \bibfield  {author} {\bibinfo {author} {\bibfnamefont {J.}~\bibnamefont
  {Bouchaud}}, \bibinfo {author} {\bibfnamefont {A.}~\bibnamefont {Comtet}},
  \bibinfo {author} {\bibfnamefont {A.}~\bibnamefont {Georges}}, \ and\
  \bibinfo {author} {\bibfnamefont {P.~L.}\ \bibnamefont {Doussal}},\ }\href
  {\doibase http://dx.doi.org/10.1016/0003-4916(90)90043-N} {\bibfield
  {journal} {\bibinfo  {journal} {Ann. Phys.}\ }\textbf {\bibinfo {volume}
  {201}},\ \bibinfo {pages} {285} (\bibinfo {year} {1990})}\BibitemShut
  {NoStop}%
\bibitem [{\citenamefont {Bouchaud}\ and\ \citenamefont
  {Georges}(1990)}]{BouchaudReview}%
  \BibitemOpen
  \bibfield  {author} {\bibinfo {author} {\bibfnamefont {J.-P.}\ \bibnamefont
  {Bouchaud}}\ and\ \bibinfo {author} {\bibfnamefont {A.}~\bibnamefont
  {Georges}},\ }\href {\doibase http://dx.doi.org/10.1016/0370-1573(90)90099-N}
  {\bibfield  {journal} {\bibinfo  {journal} {Phys. Rep.}\ }\textbf {\bibinfo
  {volume} {195}},\ \bibinfo {pages} {127} (\bibinfo {year}
  {1990})}\BibitemShut {NoStop}%
\bibitem [{\citenamefont {Nelson}\ and\ \citenamefont {Shnerb}(1998)}]{popbio}%
  \BibitemOpen
  \bibfield  {author} {\bibinfo {author} {\bibfnamefont {D.~R.}\ \bibnamefont
  {Nelson}}\ and\ \bibinfo {author} {\bibfnamefont {N.~M.}\ \bibnamefont
  {Shnerb}},\ }\href {\doibase 10.1103/PhysRevE.58.1383} {\bibfield  {journal}
  {\bibinfo  {journal} {Phys. Rev. E}\ }\textbf {\bibinfo {volume} {58}},\
  \bibinfo {pages} {1383} (\bibinfo {year} {1998})}\BibitemShut {NoStop}%
\bibitem [{\citenamefont {Dahmen}\ \emph {et~al.}(1999)\citenamefont {Dahmen},
  \citenamefont {Nelson},\ and\ \citenamefont {Shnerb}}]{popbio2}%
  \BibitemOpen
  \bibfield  {author} {\bibinfo {author} {\bibfnamefont {K.~A.}\ \bibnamefont
  {Dahmen}}, \bibinfo {author} {\bibfnamefont {D.~R.}\ \bibnamefont {Nelson}},
  \ and\ \bibinfo {author} {\bibfnamefont {N.~M.}\ \bibnamefont {Shnerb}},\
  }in\ \href@noop {} {\emph {\bibinfo {booktitle} {Statistical mechanics of
  biocomplexity}}}\ (\bibinfo  {publisher} {Springer Berlin Heidelberg},\
  \bibinfo {year} {1999})\ pp.\ \bibinfo {pages} {124--151}\BibitemShut
  {NoStop}%
\bibitem [{\citenamefont {Lubensky}\ and\ \citenamefont {Nelson}(2000)}]{DNA1}%
  \BibitemOpen
  \bibfield  {author} {\bibinfo {author} {\bibfnamefont {D.~K.}\ \bibnamefont
  {Lubensky}}\ and\ \bibinfo {author} {\bibfnamefont {D.~R.}\ \bibnamefont
  {Nelson}},\ }\href {\doibase 10.1103/PhysRevLett.85.1572} {\bibfield
  {journal} {\bibinfo  {journal} {Phys. Rev. Lett.}\ }\textbf {\bibinfo
  {volume} {85}},\ \bibinfo {pages} {1572} (\bibinfo {year}
  {2000})}\BibitemShut {NoStop}%
\bibitem [{\citenamefont {Lubensky}\ and\ \citenamefont {Nelson}(2002)}]{DNA2}%
  \BibitemOpen
  \bibfield  {author} {\bibinfo {author} {\bibfnamefont {D.~K.}\ \bibnamefont
  {Lubensky}}\ and\ \bibinfo {author} {\bibfnamefont {D.~R.}\ \bibnamefont
  {Nelson}},\ }\href {\doibase 10.1103/PhysRevE.65.031917} {\bibfield
  {journal} {\bibinfo  {journal} {Phys. Rev. E}\ }\textbf {\bibinfo {volume}
  {65}},\ \bibinfo {pages} {031917} (\bibinfo {year} {2002})}\BibitemShut
  {NoStop}%
\bibitem [{\citenamefont {Fisher}\ and\ \citenamefont
  {Kolomeisky}(1999)}]{fisher1999force}%
  \BibitemOpen
  \bibfield  {author} {\bibinfo {author} {\bibfnamefont {M.~E.}\ \bibnamefont
  {Fisher}}\ and\ \bibinfo {author} {\bibfnamefont {A.~B.}\ \bibnamefont
  {Kolomeisky}},\ }\href@noop {} {\bibfield  {journal} {\bibinfo  {journal} {P.
  Natl. Acad. Sci. USA}\ }\textbf {\bibinfo {volume} {96}},\ \bibinfo {pages}
  {6597} (\bibinfo {year} {1999})}\BibitemShut {NoStop}%
\bibitem [{\citenamefont {Rief}\ \emph {et~al.}(2000)\citenamefont {Rief},
  \citenamefont {Rock}, \citenamefont {Mehta}, \citenamefont {Mooseker},
  \citenamefont {Cheney},\ and\ \citenamefont {Spudich}}]{rief2000myosin}%
  \BibitemOpen
  \bibfield  {author} {\bibinfo {author} {\bibfnamefont {M.}~\bibnamefont
  {Rief}}, \bibinfo {author} {\bibfnamefont {R.~S.}\ \bibnamefont {Rock}},
  \bibinfo {author} {\bibfnamefont {A.~D.}\ \bibnamefont {Mehta}}, \bibinfo
  {author} {\bibfnamefont {M.~S.}\ \bibnamefont {Mooseker}}, \bibinfo {author}
  {\bibfnamefont {R.~E.}\ \bibnamefont {Cheney}}, \ and\ \bibinfo {author}
  {\bibfnamefont {J.~A.}\ \bibnamefont {Spudich}},\ }\href@noop {} {\bibfield
  {journal} {\bibinfo  {journal} {P. Natl. Acad. Sci. USA}\ }\textbf {\bibinfo
  {volume} {97}},\ \bibinfo {pages} {9482} (\bibinfo {year}
  {2000})}\BibitemShut {NoStop}%
\bibitem [{\citenamefont {Kafri}\ \emph {et~al.}(2004)\citenamefont {Kafri},
  \citenamefont {Lubensky},\ and\ \citenamefont {Nelson}}]{brm1}%
  \BibitemOpen
  \bibfield  {author} {\bibinfo {author} {\bibfnamefont {Y.}~\bibnamefont
  {Kafri}}, \bibinfo {author} {\bibfnamefont {D.~K.}\ \bibnamefont {Lubensky}},
  \ and\ \bibinfo {author} {\bibfnamefont {D.~R.}\ \bibnamefont {Nelson}},\
  }\href {\doibase http://dx.doi.org/10.1529/biophysj.103.036152} {\bibfield
  {journal} {\bibinfo  {journal} {Biophys. J.}\ }\textbf {\bibinfo {volume}
  {86}},\ \bibinfo {pages} {3373} (\bibinfo {year} {2004})}\BibitemShut
  {NoStop}%
\bibitem [{\citenamefont {Kafri}\ \emph {et~al.}(2005)\citenamefont {Kafri},
  \citenamefont {Lubensky},\ and\ \citenamefont {Nelson}}]{brm2}%
  \BibitemOpen
  \bibfield  {author} {\bibinfo {author} {\bibfnamefont {Y.}~\bibnamefont
  {Kafri}}, \bibinfo {author} {\bibfnamefont {D.~K.}\ \bibnamefont {Lubensky}},
  \ and\ \bibinfo {author} {\bibfnamefont {D.~R.}\ \bibnamefont {Nelson}},\
  }\href {\doibase 10.1103/PhysRevE.71.041906} {\bibfield  {journal} {\bibinfo
  {journal} {Phys. Rev. E}\ }\textbf {\bibinfo {volume} {71}},\ \bibinfo
  {pages} {041906} (\bibinfo {year} {2005})}\BibitemShut {NoStop}%
\bibitem [{\citenamefont {Hatano}\ and\ \citenamefont
  {Nelson}(1996)}]{Hatano1}%
  \BibitemOpen
  \bibfield  {author} {\bibinfo {author} {\bibfnamefont {N.}~\bibnamefont
  {Hatano}}\ and\ \bibinfo {author} {\bibfnamefont {D.~R.}\ \bibnamefont
  {Nelson}},\ }\href {\doibase 10.1103/PhysRevLett.77.570} {\bibfield
  {journal} {\bibinfo  {journal} {Phys. Rev. Lett.}\ }\textbf {\bibinfo
  {volume} {77}},\ \bibinfo {pages} {570} (\bibinfo {year} {1996})}\BibitemShut
  {NoStop}%
\bibitem [{\citenamefont {Hatano}\ and\ \citenamefont
  {Nelson}(1997)}]{Hatano2}%
  \BibitemOpen
  \bibfield  {author} {\bibinfo {author} {\bibfnamefont {N.}~\bibnamefont
  {Hatano}}\ and\ \bibinfo {author} {\bibfnamefont {D.~R.}\ \bibnamefont
  {Nelson}},\ }\href {\doibase 10.1103/PhysRevB.56.8651} {\bibfield  {journal}
  {\bibinfo  {journal} {Phys. Rev. B}\ }\textbf {\bibinfo {volume} {56}},\
  \bibinfo {pages} {8651} (\bibinfo {year} {1997})}\BibitemShut {NoStop}%
\bibitem [{\citenamefont {Shnerb}\ and\ \citenamefont
  {Nelson}(1998)}]{Shnerb1}%
  \BibitemOpen
  \bibfield  {author} {\bibinfo {author} {\bibfnamefont {N.~M.}\ \bibnamefont
  {Shnerb}}\ and\ \bibinfo {author} {\bibfnamefont {D.~R.}\ \bibnamefont
  {Nelson}},\ }\href {\doibase 10.1103/PhysRevLett.80.5172} {\bibfield
  {journal} {\bibinfo  {journal} {Phys. Rev. Lett.}\ }\textbf {\bibinfo
  {volume} {80}},\ \bibinfo {pages} {5172} (\bibinfo {year}
  {1998})}\BibitemShut {NoStop}%
\bibitem [{\citenamefont {Hurowitz}\ and\ \citenamefont {Cohen}(2016)}]{neg}%
  \BibitemOpen
  \bibfield  {author} {\bibinfo {author} {\bibfnamefont {D.}~\bibnamefont
  {Hurowitz}}\ and\ \bibinfo {author} {\bibfnamefont {D.}~\bibnamefont
  {Cohen}},\ }\href {\doibase 10.1038/srep22735} {\bibfield  {journal}
  {\bibinfo  {journal} {Scientific Reports}\ }\textbf {\bibinfo {volume} {6}},\
  \bibinfo {pages} {22735} (\bibinfo {year} {2016})}\BibitemShut {NoStop}%
\bibitem [{\citenamefont {Hurowitz}\ \emph {et~al.}(2013)\citenamefont
  {Hurowitz}, \citenamefont {Rahav},\ and\ \citenamefont {Cohen}}]{nef}%
  \BibitemOpen
  \bibfield  {author} {\bibinfo {author} {\bibfnamefont {D.}~\bibnamefont
  {Hurowitz}}, \bibinfo {author} {\bibfnamefont {S.}~\bibnamefont {Rahav}}, \
  and\ \bibinfo {author} {\bibfnamefont {D.}~\bibnamefont {Cohen}},\ }\href
  {\doibase 10.1103/PhysRevE.88.062141} {\bibfield  {journal} {\bibinfo
  {journal} {Phys. Rev. E}\ }\textbf {\bibinfo {volume} {88}},\ \bibinfo
  {pages} {062141} (\bibinfo {year} {2013})}\BibitemShut {NoStop}%
\bibitem [{\citenamefont {Hurowitz}\ and\ \citenamefont {Cohen}(2014)}]{nes}%
  \BibitemOpen
  \bibfield  {author} {\bibinfo {author} {\bibfnamefont {D.}~\bibnamefont
  {Hurowitz}}\ and\ \bibinfo {author} {\bibfnamefont {D.}~\bibnamefont
  {Cohen}},\ }\href {\doibase 10.1103/PhysRevE.90.032129} {\bibfield  {journal}
  {\bibinfo  {journal} {Phys. Rev. E}\ }\textbf {\bibinfo {volume} {90}},\
  \bibinfo {pages} {032129} (\bibinfo {year} {2014})}\BibitemShut {NoStop}%
\bibitem [{\citenamefont {Molinari}(2008)}]{det1}%
  \BibitemOpen
  \bibfield  {author} {\bibinfo {author} {\bibfnamefont {L.~G.}\ \bibnamefont
  {Molinari}},\ }\href {\doibase http://dx.doi.org/10.1016/j.laa.2008.06.015}
  {\bibfield  {journal} {\bibinfo  {journal} {Linear Algebra Appl.}\ }\textbf
  {\bibinfo {volume} {429}},\ \bibinfo {pages} {2221} (\bibinfo {year}
  {2008})}\BibitemShut {NoStop}%
\end{thebibliography}%

%

\end{document}